# Complete Field Characterization of Ultrashort Pulses in Fiber Photonics

Elena A. Anashkina, Alexey V. Andrianov, Maxim Yu. Koptev, and Arkady V. Kim

*Abstract*—We report a simple fiber-implemented technique for complete reconstruction of intensity profile and phase of ultrashort laser pulses based on processing only pulse spectrum and two self-phase modulated spectra measured after a short piece of optical fiber. Its applicability is shown on an example of a fiber optical system in the telecommunication range. A retrieval algorithm in a dispersionless approximation and with considering dispersion effects is developed. The obtained results are confirmed by independent measurements using the second-harmonic generation frequency-resolved optical gating technique and by reconstructing purposely introduced signal features. We also provide estimates demonstrating great opportunities for implementing this technique in all-waveguide optical systems ranging from optical communications to nanophotonics with femtojoule pulses as well as to mid-IR photonics, where specialty fibers with huge optical nonlinearities can be used.

## I. INTRODUCTION

ULTRASHORT optical pulses are playing an increasing role in societal applications, including telecommunications, medicine, and technology [1,2]. This attaches great importance to the development of simple, robust and cost-effective methods of pulse characterization, especially for ultrafast photonics dealing with very low energy pulses. It should be noted that free-space techniques of pulse characterization are now sufficiently well established. We particularly note the widely used phase-sensitive techniques such as SHG-FROG (second-harmonic generation frequency-resolved optical gating) [3] and SPIDER (spectral phase interferometry for direct electric-field reconstruction) [4]. Their numerous modifications as well original methods with different advantages and benefits for particular targets have also been proposed [5-14] to avoid some drawbacks of the above methods, such as the presence of ambiguities, insufficient temporal and/or spectral resolution and range, inherent inability to completely reconstruct pulse intensity and/or phase, inability to measure complex signals, and misleading results due to loss of information during averaging over many pulses. The advanced methods allow continuous single-shot measuring of complex signals including optical rogue waves and stochastic pulses [15-18]. Some of the developed free-space techniques are now realized in waveguide compatible photonic devices [19-23]. However, we believe that this is an initial stage of developing optical metrology in ultrafast fiber photonics not only because of low energy of pulses but also because of the need to use waveguide compatible optical elements.

In this work, we would like to draw the photonics community attention to a new platform of complete pulse reconstruction based on measurements of self-phase modulated spectra in nonlinear fibers and waveguides that are compatible with optical devices and thus provide the all-waveguide format. It should be emphasized that the great advantage of this method realized with free-space techniques is the use of a very thin Kerr plate for which dispersion effects are negligible [24]. In fact, the method and algorithm proposed were realized for reconstruction of TW and PW single-shot laser pulses based on spectral measurements of two self-phase modulated (SPM) spectra after thin plastic films with Kerr nonlinearity which requires a sufficiently broad and uniform laser beam and neglecting other linear and nonlinear effects [24]. The technique for characterization of ultrashort laser pulses based on the fundamental and one SPM-spectrum in the dispersionless approximation was earlier developed in [25, 26]. But the mentioned technique is ambiguous and requires knowledge of nonlinearity and peak pulse power [25]. These challenges may be overcome using two SPM-spectra. However, in the case of low energy pulses in fiber photonics, dispersion effects which are important during pulse propagation over extended nonlinear medium should be included in the field reconstruction procedure.

Here we report a novel experimental low-energy fiber implementation and numerical algorithm taking into account dispersion effects for reconstruction of intensity profile and phase of ultrashort pulses without temporal ambiguity using only a short piece of fiber, a spectrometer, and a power meter. The experimental and numerical study for silica fibers and theoretical estimations for soft-glass fibers and nanowires show that this very simple technique may be used for

Manuscript received May 12, 2017. Experimental implementation of the method as well as computer realization of the novel retrieval algorithm were supported by the Russian Science Foundation (project №16-12-10472); E.A.Anashkina also obtained support from the Russian Foundation for Basic Research for theoretical study of the method and its applicability, including for mid-IR pulses (grant №16-32-60053).

The authors are with the Institute of Applied Physics of the Russian Academy of Sciences, Nizhny Novgorod 603950, Russia. (e-mail: elena.anashkina@gmail.com; alex.v.andrianov@gmail.com; max-koptev@ya.ru; kim@ufp.appl.sci-nnov.ru)

complete characterization of pulses generated by a wide class of ultrashort photonic sources by exploiting different fibers (or other waveguides) with Kerr nonlinearity: silica fibers for 1-2 μm wavelengths and energies of order 1 nJ and higher, soft-glass fibers [27] for near- and mid-IR and energies down to ~1 pJ, nanowires for extremely low-energy optical devices down to sub-pJ or even ten fJ. The technique has been examined on various ultrashort pulses generated by a home-made Er-doped fiber laser source. The obtained results are confirmed by independent SHG-FROG-measurements and by reconstruction of specially prepared pulse features.

## II. EXPERIMENTAL SETUP

The schematic of the experiment is shown in Fig. 1.

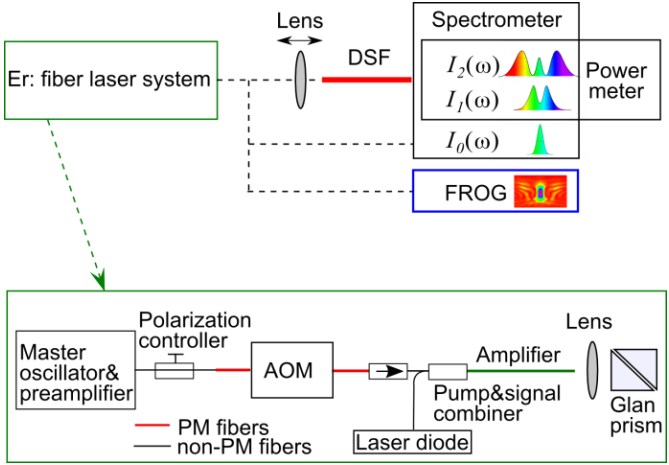

Fig. 1. Schematic of the experiment.

We use an Er-doped fiber laser system as a signal source. The system consists of a master oscillator operating at 1.57 μm with a pulse repetition rate of 50 MHz, an Er-doped preamplifier, a polarization controller, a polarization-maintaining (PM) acousto-optic modulator (AOM) used to decrease the pulse repetition rate, a Faraday isolator and Er-doped fiber amplifier. The diode-pumped master oscillator at 975 nm is passively mode-locked by nonlinear rotation of the polarization ellipse of a femtosecond pulse due to the optical Kerr effect. The resulting pulse duration is of order 100 fs, the pulse energy is about nJ level, and the average power is up to tens of mW for a repetition rate of 10 MHz. Note that a Raman soliton may be formed in the amplifier during propagation. In this case, a small part of the signal remains at the initial central wavelength and is time-spaced with respect to the soliton at the amplifier output.

Part of the scheme was based on PM-fibers (AOM and its pigtails). Polarization at the input of the PM fiber pigtail can be adjusted by a polarization controller. The pulses entering the PM pigtail split along both fast and slow axes of the fiber. Two orthogonally polarized pulse replicas propagate simultaneously and accumulate different time delays due to the group velocity difference between two different polarized components. At the output of the amplifier the signal propagates through a Glan prism to ensure linear polarization. At a certain orientation of the Glan prism, both orthogonally polarized pulse replicas contribute to the output signal, which results in a double pulse at the output. We purposely prepared double pulses to test the proposed reconstruction technique, whereas for other tests we chose only one polarization.

The signal is launched by means of a focusing D-ZK3 glass lens into a 2-cm piece of a single-mode dispersion-shifted fiber (DSF). The dispersion of the lens is insignificant here. DSF has a zero-dispersion wavelength (ZDW) of about 1.45 μm, anomalous dispersion of about $\beta_2 = -20$ ps²/km at 1.57 μm, and the estimated nonlinear coefficient $\gamma = 4$ (W km)⁻¹. The estimations show that for pulses with energy of about 1 nJ and less at the initial stage of propagation in DSF, the dynamics is determined by a Kerr nonlinearity, because Raman scattering, dispersion, optical loss, and self-steepening may be neglected. But with increasing input energy or fiber length, when a broadened spectral wing approaches the ZDW, dispersion effects should be considered. We have studied both cases when dispersion of DSF can be neglected and when its impact gives notable corrections. The pulse spectrum before DSF and two SPM-spectra at the output of DSF corresponding to various output average powers have been recorded. To reduce average power, the lens focus has been tuned off the optimal location. This way of energy decrease at the fundamental is very simple and allows preserving the temporal and spectral structure of the input signal.

The SHG FROG home-built apparatus based on a 100 μm BBO nonlinear crystal and a spectrometer is also used to characterize optical pulses in the time domain for independent field reconstruction measurements.

## III. NUMERICAL ALGORITHM

The numerical algorithm of field characterization based on measuring only three spectra was first described in [24]. The first spectrum $I_0(\omega)$ is the fundamental (input pulse spectrum), the second $I_1(\omega)$ and the third $I_2(\omega)$ ones are after a thin Kerr plate, where $\omega$ is the angular frequency counted from central frequency $\omega_0$. But in this case relevant to free-space optics experiments dealing with high enough pulse energies, nonlinear self-phase modulation needed for measurements can be acquired using only very thin plates for which dispersion is negligible. However, this is not so if instead of a thin Kerr plate a piece of nonlinear waveguide (fiber) is used. Here we generalize the numerical algorithm to an extended nonlinear medium allowing for dispersion effects in the pulse propagation model. Note that the $I_1(\omega)$ and $I_2(\omega)$ spectra are recorded for different pulse energies $W_1$ and $W_2$ with known ratio $\eta = W_2 / W_1$ (or by using two pieces of fiber with different lengths). The pulse energy is

$$W_j = \int_{-\infty}^{\infty} |E_j(t)|^2 \, dt \,, \tag{1}$$

where $t$ is the time in the retarded frame and $E_j(t)$ is the slowly varying pulse envelope [28] of the signal designated by $j$. The Fourier transformed complex amplitude $\tilde{E}_j(\omega)$ is given by

$$\tilde{E}_j(\omega) = \int_{-\infty}^{\infty} E_j(t)\exp(i\omega t)dt.$$ (2)

A fast Fourier transform (FFT) is employed to re-calculate functions from the time to the frequency domain and an inverse FFT (IFFT) for re-calculation from the frequency to the time domain. If one can neglect Raman nonlinearity, optical losses, and self-steepening, the dimensionless pulse envelopes $E_1(t)$ and $E_2(t)$ after propagating through a short piece of fiber can be simulated from the nonlinear Schrödinger equation by the split-step Fourier method (SSFM) [28]:

$$\frac{\partial E_j(z,t)}{\partial z} = i\sum_{q=2}^{M}\frac{(i)^q}{k!}\beta^{(q)}\frac{\partial^q E_j(z,t)}{\partial t^q},$$
$$+ i\gamma P_j\left|E_j(z,t)\right|^2 E_j(z,t)$$ (3)

where $z$ is the distance along the fiber, $\gamma$ is fiber nonlinear coefficient [28], $\beta$ is the wavenumber of the fundamental mode, and $P_j$ is the input peak power proportional to $W_j$, $M$ is the maximum dispersion order considered. It should be emphasized that the main requirement for the proposed method and algorithm is that the evolution equation, in our case Eq. (3), should describe pulse propagation deterministically. This means, firstly, that the method in the frame of Eq. (3) allows reliable operation with normal dispersion fibers; however, fibers with anomalous dispersion can also be used if the pulse propagation mode is free of modulation instability. Secondly, Eq. (3) can be generalized by including Raman nonlinearity and/or other effects such as nonlinear dispersion, frequency dependence of mode field, waveguide loss [28] and so on for coherent pulse propagation, which allows extending the applicability conditions for a wide class of photonic devices. For practical applications, if a pulse spectrum lies far enough from the zero-dispersion wavelength, only quadratic dispersion ($M = 2$) may be considered. However, if a spectral wing comes to near zero dispersion wavelength, cubic dispersion ($M = 3$) should also be taken into consideration.

For the characterized pulse $\tilde{E}_0(\omega) = [I_0(\omega)]^{1/2}\exp[i\,\varphi_0(\omega)]$ it is desirable to choose a spectral phase $\varphi_0(\omega)$ minimizing the difference (error $\Delta$) between two measured SPM-spectra and the corresponding simulation. Error is defined by

$$\Delta = \left\{\sum_{k=1}^{N}\left[I_1(\omega_k) - |\tilde{E}_1(\omega_k)|^2\right]^2\right.$$
$$\left. + \sum_{k=1}^{N}\left[I_2(\omega_k) - |\tilde{E}_2(\omega_k)|^2\right]^2\right\},$$ (4)

where $N$ is the number of points in the frequency domains designated by $k$.

The error is minimized using the scheme shown in Fig. 2. If $\eta = 2$, in the dispersionless approximation, the time-equivalent Gerchberg–Saxton like algorithm without simulation of pulse backward propagation can be applied using in parallel multi-SPM-spectra information [24].

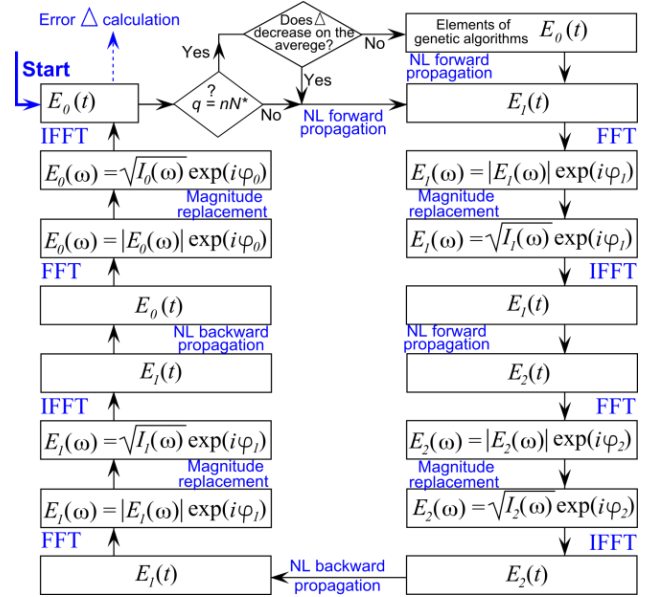

Fig. 2. Scheme of the numerical algorithm. Here $q$ is a number of current iteration, $n$ is a natural number, $N^*$ =const (usually ~ 30 – 100).

However, here we demonstrate an alternative procedure based on Gerchberg–Saxton algorithm [29] by sequentially applying multi-step only spectral intensity measurements. It may be very convenient for fiber implementation because it does not require keeping precise energies ratio of 2 and allows easy incorporation of the dispersion impact for simulating forward and backward pulse propagation [28]. When choosing fibers, it is important to have notably different original and transformed spectra. In this case, maximum phase shift (or B-integral) in the fiber should be of order 1. Allowance for dispersion significantly increases computation time. It may be roughly estimated as follows. In our implementation, the program contains 10 FFT per iteration (shown in Fig. 2) in the dispersionless approximation and $(2 + 8\cdot N_z)$ FFT with dispersion taken into account, where $N_z$ is the number of segments into which the fiber is divided to apply the SSFM method. It should be expected that with other equal parameters, the time increases by approximately $N_z$ times, which is confirmed by direct runs of the program for different $N_z$.

Note that using a fundamental spectrum and one transformed spectrum does not uniquely specify unknown phase. Consider as an example two time-symmetric pulses with the same amplitude distributions of the fields $E_{01}(t) = E_{01}(-t)$, $|E_{02}(t)| = |E_{01}(t)|$ and phases proportional to the intensity distribution, but differing in sign $\psi_{01,02}(t) = \pm b|E_{01}(t)|^2$, where $b = const$. Their spectra are identical. The transformed signals are $E_{1,2}(t) = |E_{01}(t)| \cdot \exp[i|E_{01}(t)|^2(\pm b + \gamma PL)]$. The exact value of $\gamma PL$ is not known as a rule. Note that, if $(\gamma PL)_1$ is chosen for the pulse $E_{01}$ and $(\gamma PL)_2 = (\gamma PL)_1 + 2b$ for the pulse $E_{02}$, then the transformed spectra are identical too. This means that in practice, for sufficiently symmetric pulses with an almost quadratic phase, it may be difficult to correctly determine the sign of the chirp. Two output spectra used together with the input one exclude ambiguity.

Starting with the guessed spectral phase $\varphi_0(\omega)$, $E_0(t)$ is constructed numerically, and nonlinear propagation for energy $W_1$ is simulated by numerical integration of Eq. (3). After FFT of $E_1(t)$ resulting in $\tilde{E}_1(\omega)$ the spectral phase $\varphi_1(\omega)$ remains unchanged but its magnitude is replaced by the experimentally measured square roots from the SPM-spectra $I_1(\omega)$, and IFFT is executed. Further, $E_2(t)$ is calculated using energy $W_2$, FFT is executed, spectral phase $\varphi_2(\omega)$ is also saved but the magnitude is replaced by square roots of $I_2(\omega)$. Next, an analogous procedure for back propagation is simulated (replacing $z \rightarrow -z$, $t \rightarrow -t$ in Eq. (3)) and the cycle is completed by replacing the fundamental spectrum by the experimental data, saving the spectral phase with subsequent IFFT to obtain $E_0(t)$. The error is estimated at each iteration from formula (4). The procedure of saving the phase replacing the amplitude is standard for the Gerchberg–Saxton algorithm and its numerous variations and modifications. To improve convergence, elements of the genetic algorithm such as crossover and mutation [30] are used, if the error does not decrease on the average for a large enough number of iterations $N^*$ (usually $N^* \sim 30 - 100$). Thus, we use elements of the genetic algorithm to avoid local minimum and to converge to the global one. However, we cannot rigorously state that the program must always find a global minimum. The algorithm requires setting the value $\gamma P_j$ which is not well-known as a rule. We have run the algorithm for various $\gamma P_j$ (exhaustive search with a constant step d$\gamma P_j$ in the pre-estimated range of interest) and selected the value giving a minimum $\Delta_{min}$ like in [24]. It was earlier demonstrated by simulated examples that the minimum of the function $\Delta_{min}(\gamma P_j)$ is sharp enough [24]. We believe that a reconstruction is reliable if error $\Delta_{min}$ is relatively small (the retrieved SPM spectra are similar to the measured ones). It is stable in the considered case. If there is a significant divergence between the retrieved and measured spectra, the found field distribution is not reliable.

## IV. RESULTS AND DISCUSSION

To begin with, we have characterized signals directly from the fiber source ("Test I", Fig. 3(a-e)) and with a 6.5 cm-long bulk glass placed after the source before DSF ("Test II", Fig. 3(f-j)) to add a certain chirp and retrieve it using the proposed technique.

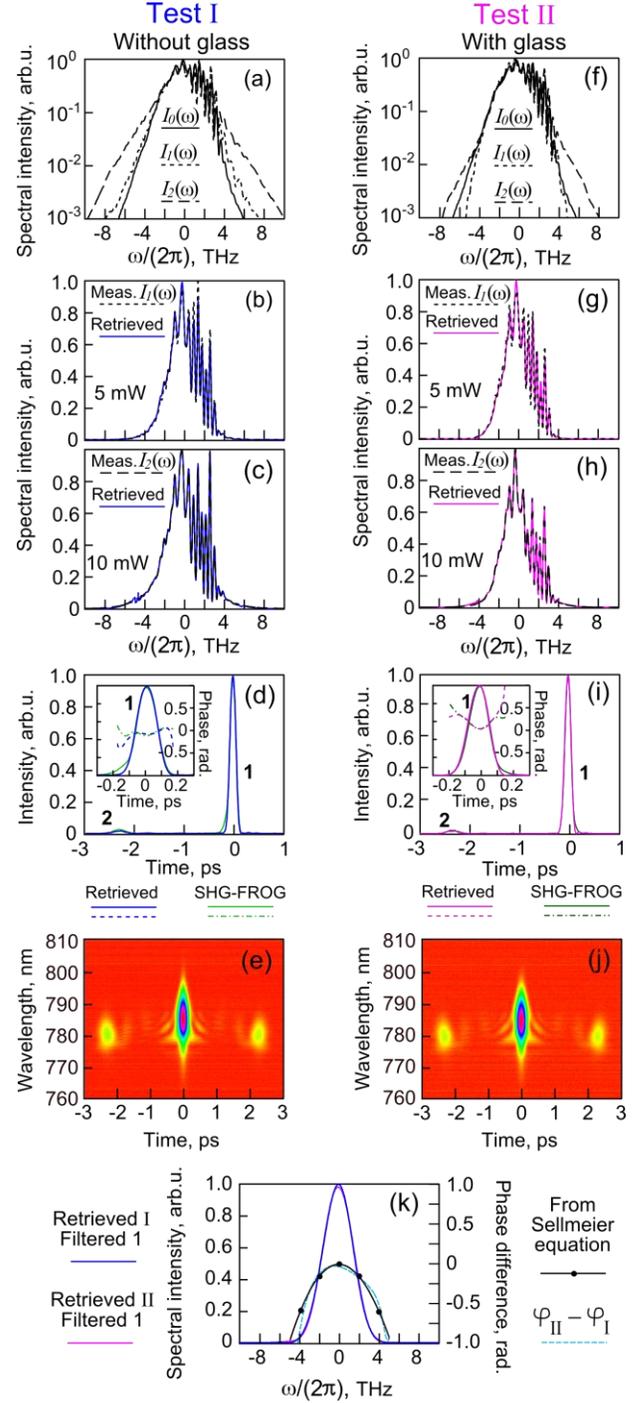

Fig. 3. Pulse reconstruction of unchirped and purposely chirped pulses in the dispersionless approximation. Experimental signals: directly from the output of Er: fiber laser source ("Test I") and after propagating through 6.5 cm-thick glass ("Test II"). Their measured fundamental and SPM spectra (a) and (f); measured and retrieved SPM spectra for various average power (b, c) and (g, h); reconstructed temporal intensity profiles and phases with the proposed technique and with SHG-FROG (d) and (i); measured SHG-FROG-traces (e) and (j). Spectra of signals 1 filtered in the time domain for "Test I" and "Test II" and difference of their spectral phases obtained from reconstruction and calculated by the Sellmeier equation for a 6.5 cm glass (k).

The algorithm has been run in the dispersionless approximation for the fundamental and two SPM spectra for average output powers of 5 and 10 mW (Fig. 3(a) for "Test I" and Fig. 3(f) for "Test II"). One can see a perfect agreement between the retrieved and the original SPM spectra (Fig. 3(b, c) for "Test I" and (g, h) for "Test II"). Fringes in the spectra correspond to the interference between pulse "1" having a duration of 140 fs formed in the Er: fiber amplifier and the remaining part of signal "2". We believe that pulse "1" downshifts its frequency due to the Raman effect in the amplifier itself and is located at the trailing edge of the non-converted part "2" of the signal [28]. The reconstructed temporal intensity distributions in comparison with the FROG-retrieved ones are shown in Fig. 3(d) for "Test I" and in Fig. 3(i) for "Test II". Pulses "1" and "2" are labeled in Fig. 3(d,i). SHG-FROG traces [3] depicted in Fig. 3 (e) and (j). So, the proposed technique and the FROG give similar results. The contributions to the temporal phase for "Test II" of the bulk glass dispersion and of SPM have opposite signs, so one can see spectrum narrowing for $I_1(\omega)$ in Fig. 3(f) due to phase flattening. But for a higher pulse energy corresponding to $I_2(\omega)$, an additional phase from SPM dominates and the spectrum is broadened.

Further, we filtered the pulses for "Test I" and "Test II" centered near t = 0 and applied to them FFT. Their spectra are shown in Fig. 3(k). The spectral phase difference ($\varphi_{II} - \varphi_I$) between the pulses chirped by glass and unchirped pulses is also shown in Fig. 3(k). As the phase difference originates from the bulk glass, it has been easily calculated using the Sellmeier equation [28]. So, Fig. 3(k) demonstrates a very good agreement between the reconstructed and the theoretically estimated phase difference.

The following "Test III" concerns double pulses, which are challenging for many techniques. We have purposely prepared such pulses by detuning the polarization controller and PM AOM. The polarization at the input of the AOM PM fiber pigtail can be varied so that pulses corresponding to fast and slow axes propagating along the fiber result in pulse doubling at the output because of their different group velocities. The algorithm has been run for the fundamental and two SPM spectra for output powers of 7.5 and 20 mW (see Fig. 4(a)).

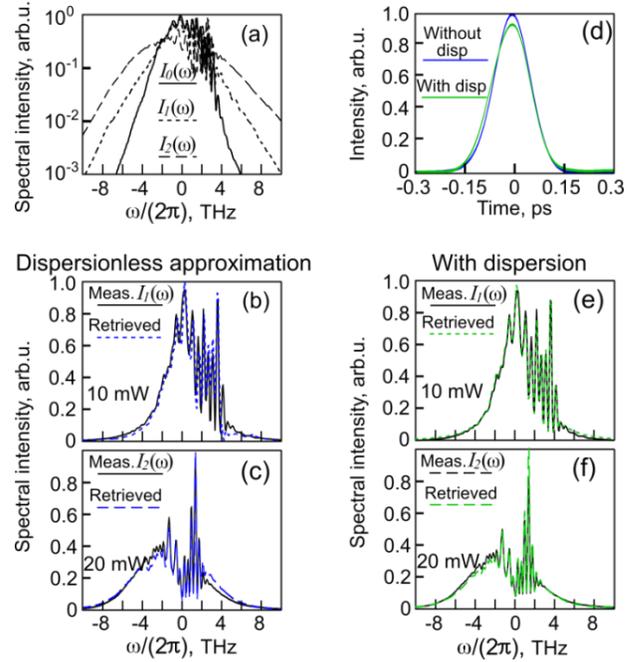

Fig. 5. Pulse reconstruction in both the dispersionless approximation and with dispersion taken into account. Measured fundamental and SPM spectra (a); measured and retrieved SPM spectra in the dispersionless approximation (b, c) and with consideration of dispersion impact (d); reconstructed temporal intensity profiles and phases with and without allowance for dispersion.

Here we have specially chosen energy ratio $\eta$ differing from 2 to demonstrate algorithm applicability for this case. The retrieved SPM-spectra are in a very good agreement with the experimental data (see Fig. 4(b,c)). A temporal intensity profile is confirmed by SHG-FROG measurements, the results of which are shown in Fig. 4(d). FROG-trace is shown in Fig. 4(e). As pulses 1 and 2 are two replicas of one, their normalized intensity profiles and phases should coincide. This statement is illustrated in Fig. 4(f) for the retrieved signal.

We see that for the demonstrated "Test I-III", the impact of dispersion is negligible, as the measured and calculated SPM-spectra are in a very good agreement. This means that for particular cases we can choose fiber and input pulse parameters such that the dispersion effects should be small enough for the proposed method of field reconstruction. In general, however, the dispersion effects in the pulse propagation over extended media should be included in our reconstruction scheme. We applied the developed method to "Test IV" and showed that allowance for dispersion effects actually gives notable corrections in the retrieved field and improves similarity of experimental and reconstructed SPM-spectra. For "Test IV" the average power at the output of a piece of fiber has been increased up to 20 mW. Modeling shows that this leads to pulse shortening at the fiber output associated with the initial stage of higher-order soliton compression [28]. Besides, the higher frequency spectral wing approaches ZDW, and third-order dispersion also impacts on spectral shape. The algorithm has been run for $\beta_2 = -20$ ps$^2$/km and $\beta_3 = 0.2$ ps$^3$/km and, as is seen in Fig. 5 (a-e), gave a better agreement of SPM measured and retrieved spectra in comparison with the dispersionless approximation.

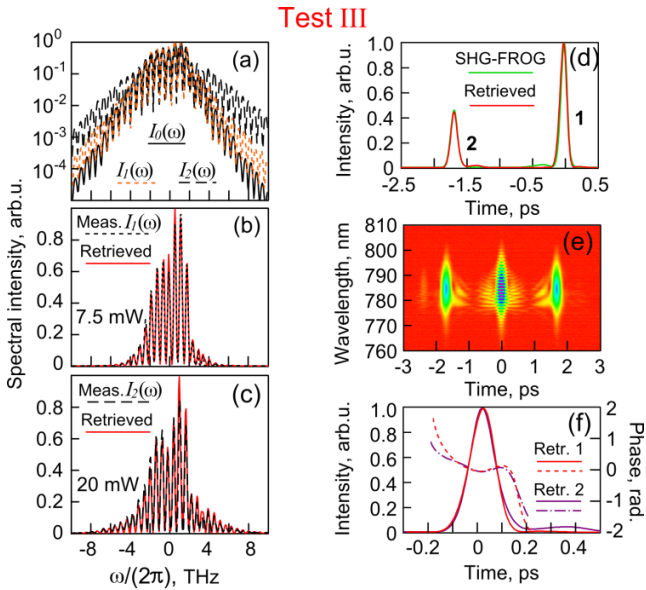

Fig. 4. Reconstruction of double-pulses. Experimental double pulse: measured fundamental and SPM spectra (a), measured and retrieved SPM spectra for 7.5 and 20 mW (b, c); reconstructed temporal intensity profile with the proposed technique and with SHG-FROG (d); measured SHG-FROG-trace (e); normalized and overlapping pulses 1 and 2 and their phases (f).

The method can be used for single-shot and for low repetition rate pulses that are hard to measure by SHG-based methods due to very low average power. The method has limitations originating from the properties of the Fourier transform. So, the temporal range is determined by spectrometer resolution and temporal resolution depends on full spectral width. In our opinion, this technique is not good for few-cycle pulses because of precise knowledge of fiber parameters required for correct simulation of pulse propagation, but we did not check this experimentally. We believe that for measurements of pulses with duration starting from tens of fs and peak powers of about tens of kW it would be very handy to use a piece of silica fiber for the wavelengths of 1-2 μm. But specially designed fibers can significantly extend the range of signal parameters. For example, commercially available step-index ZBLAN-fibers [31] may allow characterizing ultrashort nJ-pulses in the mid-IR (up to ~4μm). ZBLAN fibers have nonlinearity similar to silica, admissible anomalous dispersion and low optical losses. This is especially relevant for rapidly developing laser systems based on rare-earth ion doped fiber master oscillators operating near 2.8 μm [32]. We estimate that for this spectral range γ is of order 1 (W km)$^{-1}$ for the fibers parameters from [31] and $n_2 = 2.7 \cdot 10^{-16}$ cm$^2$/W from [33]. Their dispersion coefficient $\beta_2$ has an absolute value about a few tens of ps$^2$/km for core diameters of 8-10 μm [34] and can be smaller for thinner cores. Highly nonlinear microstructured tellurite fibers may allow measuring pulses with lower energy, down to tens of pJ, because their nonlinear coefficients are two orders of magnitude higher and dispersion can be controlled [35-37]. Chalcogenide fibers have even a greater nonlinearity and a broader mid-IR transparency region, which can be potentially exploited for pulse characterization up to ~15 μm [38]. The method can also be implemented for extensively developing integrable and energy-efficient photonic circuitry needed for miniaturization of optical devices [39]. Sophisticated methods based on nonlinear effects in semiconductor nanostructures have been reported for such schemes [19-23]. But we believe that the proposed technique using nanowires can significantly simplify pulse characterization. For instance, the third-order nonlinearity of silicon nanowires can exceed the nonlinearity of standard silica fibers by five orders of magnitude [40,41], which is suitable for reconstruction of pulses down to a subpicojoule or even a ten-femtojoule level (which is up to five orders of magnitude lower than the energy of order 1 nJ experimentally demonstrated here).

## V. CONCLUSIONS

In conclusion, a very simple fiber-implemented technique for reconstructing intensity profile and phase of ultrashort pulses is presented. It requires only a short piece of optical fiber, a spectrometer with a power meter, and a computer for data processing. A Gerchberg–Saxton-like algorithm has been developed by sequentially applying experimental data in three spectral intensity measurements: fundamental spectrum and two SPM-spectra after propagation through a piece of fiber for different average powers. The algorithm can operate taking into account the dispersion effects during pulse propagation along the fiber as well as the dispersionless approximation in a simpler version. The method allows operating with normal as well as anomalous waveguide dispersion. The method has no ambiguity related to time direction because SMP-spectra for $E(t)$ and $E(-t)$ are different, in contrast to SHG-FROG or SHG-autocorrelation when recorded traces for $E(t)$ and $E(-t)$ are indistinguishable. However, not all methods using SHG have the mentioned ambiguity. The d-scan technique is also free of it, even for SHG implementation [13,14].

To demonstrate that the technique is a powerful tool for complete characterization of different ultrashort optical signals, we have performed measurements with an Er: fiber laser system operating near 1.57 μm. The examined signals have included pulses with a duration of 140 fs directly from the laser output, pulses chirped by a bulk glass, and double pulses. The results demonstrated are in a good agreement with independent SHG-FROG measurements. The reconstructed certain features such as additional chirp due to bulk glass and similarity of two replicas are confirmed by theoretical calculations.

The estimations show great opportunities for implementing this technique in all-waveguide optical systems ranging from optical communications to nanophotonics with femtojoule pulses as well as to mid-IR photonics, where specialty fibers based on fluoride, tellurite and chalcogenide glasses with huge third-order nonlinearities can be used.

Thus, we believe that the technique will have a widespread photonics application primarily for fiber laser sources and all-waveguide optical systems because it is robust, low cost, allowing operation. with very low energies, and has no ambiguity related to time direction.

**Elena A. Anashkina** received the B.S. and M.S. degrees in physics from the Lobachevsky State University of Nizhni Novgorod (UNN), Nizhny Novgorod, Russia, and the Ph.D. degree in physics and mathematics from the Institute of Applied Physics of the Russian Academy of Sciences (IAP RAS), Nizhny Novgorod, Russia.

She is currently a Senior Researcher at IAP RAS. Her main field of scientific includes nonlinear optics and fiber lasers.

**Alexey V. Andrianov** received the B.S. and M.S. degrees in physics from UNN, Russia, and the Ph.D. degree in physics and mathematics from IAP RAS.

He is currently a Senior Researcher of IAP RAS. His main field of scientific interests includes fiber lasers and nonlinear optics.

**Maxim Yu. Koptev** received the B.S. and M.S. degrees in physics from UNN, Russia.

He is currently working toward the post graduate degree at IAP RAS. His main field of scientific interest is fiber lasers.

**Arkady V. Kim** received the graduate degree from the University of Nizhny Novgorod, Nizhny Novgorod, Russia, in 1974 and received the Ph.D. degree in plasma physics from the Institute of Applied Physics, Russian Academy of Sciences (IAP RAS), Nizhny Novgorod, Russia, in 1983.

He started his career at the Radio Physical Research Institute in 1974 studying microwave plasma physics and joined IAP RAS in 1977, where he is currently Head of the Extreme Nonlinear Optics laboratory. His research interests include nonlinear optics, high field physics, and high intensity laser-matter interactions.